\documentclass[final]{svjour3}
\usepackage[dvipdfmx]{graphicx}
\usepackage{graphicx}
\usepackage{rotating}
\usepackage{amssymb}
\usepackage{mathptmx}
\usepackage[numbers]{natbib}
\usepackage{subfig}
\usepackage[LGRgreek]{mathastext}

\makeatletter
\journalname{Journal of Low Temperature Physics}

\bibpunct{}{}{,}{s}{}{,}

\begin{document}

\newcommand{\hdblarrow}{H\makebox[0.9ex][l]{$\downdownarrows$}-}
\title{Characterization of high aspect ratio TiAu TES X-ray microcalorimeters array using the X-IFU Frequency Domain Multiplexing readout}

\author{E. Taralli \and L. Gottardi \and K. Nagayoshi \and M. Ridder \and S. Visser \and P. Khosropanah \and H. Akamatsu \and J. van der Kuur \and M. Bruijn \and J.-R. Gao}

\institute{E. Taralli \and L. Gottardi \and K. Nagayoshi \and M. Ridder \and S. Visser \and P. Khosropanah \and H. Akamatsu \and J. van der Kuur \and M. Bruijn \and J.R. Gao\\
SRON Netherlands Institute for Space Research\\Sorbonnelaan 2, Utrecht, 3584 CA, The Netherlands\\
\email{e.taralli@sron.nl}\\
\\
J.-R. Gao\\
Faculty of Applied Science, Delft University of Technology\\
Lorentzweg 1, 2628 CJ, Delft, The Netherlands\\
}

\maketitle

\begin{abstract}

We are developing X-ray microcalorimeters as a backup option for the baseline detectors in the X-IFU instrument on board the ATHENA space mission led by ESA and to be launched in the early 2030s.\\
5$\times$5  mixed arrays with TiAu transition-edge sensor (TES), which have different high aspect ratios and thus high resistances, have been designed and fabricated to meet the energy resolution requirement of the X-IFU instrument. Such arrays can also be used  to optimise the performance of the Frequency Domain Multiplexing (FDM) readout and lead to the final steps for the fabrication of  a large detector array.\\
In this work we present the experimental results from tens of the devices with an aspect ratio (length-to-width) ranging from 1-to-1 up to 6-to-1, measured in a single-pixel mode with a FDM readout system developed at SRON/VTT. We observed a nominal energy resolution of about 2.5 eV at 5.9~keV at bias frequencies ranging from 1 to 5~MHz. 
These detectors are proving to be the best TES microcalorimeters ever reported in Europe, being able to meet not only the requirements of the X-IFU instrument, but also those of other future challenging X-ray space missions, fundamental physics experiments, plasma characterization and material analysis.\\


\end{abstract}

\section{Introduction}

ATHENA \cite{athena} - Advanced Telescope for High-ENergy Astrophysics - is the second large mission of ESA's Cosmic Vision-programme to study astrophysical phenomena near black holes and galaxy clusters. The X-ray Integral Field Unit (X-IFU) \cite{xifu}, one of the two instruments on board, will deliver spatially-resolved high-resolution X-ray spectroscopy over a limited field of view. X-IFU requires detectors in an array of nearly 3264 pixels, being sensitive in the 0.2--12~keV energy range with 2.5~eV FWHM energy resolution below 7~keV. SRON is currently developing a Frequency Domain Multiplexing (FDM) readout system as the baseline and X-ray microcalorimeters arrays based on TiAu transition-edge sensors (TESs)  as a backup technology for the US MoAu TES array, currently being developed at NASA Goddard Space Flight Center\cite{goddard}.\\
In the last few years, a lot of progress has been reported in both the FDM readout \cite{hiroki} and in our understanding of TES physics under AC bias. The latter has been moved forward continuously by improving the detector design and performance. For instance, bare TESs have been found to be preferable to TESs with normal metal structures. It has also been shown \cite{kazu} that AC loss is bias-frequency dependent and depends also on the amount of the normal metal used in the TES. Moreover, bare TESs have shown smaller weak-link effects and the behaviour of $\alpha$ and $\beta$ are not affected by kinks \cite{nick}~\cite{taralli} in the IV curve. 
The normal resistance of the TES, together with its saturation power, plays a crucial role  in the reduction of the Josephson effect \cite{gottardi} which can affect the performance of a TES microcalorimeter under AC bias. The Josephson effects are non-dissipative so that they do not add noise to the detector, but they introduce undesired, bias-dependent nonlinearities, which limit the optimal bias range. To simplify the FDM read-out of a large TES array, these effects should be minimised. It was recently been shown that the degradation of performance from the frequency multiplexing scheme can be mitigated by the use of TESs with high resistance\cite{gottardi, kazu}. It is also important to have a thicker bilayer to reduce the internal thermal fluctuation noise.\\
All of this new knowledge has led the design and fabrication of the new SRON  5$\times$5  TES mixed arrays, containing different pixel designs, as well as the kilo-pixel uniform arrays with a single TES design at SRON. A thicker TiAu bilayer with respect to the previous SRON devices \cite{pourya} is introduced to reduce the internal thermal fluctuation noise. Furthermore, higher aspect ratios increase the normal resistance of the single pixel and thus increase the saturation power; the absence of normal metal structures reduces the weak-link effect; and a thinner membrane is used to lower the thermal conductance.\\
In this paper we present our experimental results in terms of energy resolution for tens of TiAu TESs grouped into ten different types of device with aspect ratio (length-to-width) ranging from 1-to-1 up to 6-to-1. They were measured in a single-pixel mode with bias frequencies between 1 and 5~MHz,using an 18-channel FDM readout system which is a a prototype of the FDM readout system for X-IFU.\\
\vspace{-1.0cm}
\section{TES Detector Arrays and Experimental Setup}

\begin{figure}[htbp]
\begin{center}
	\subfloat[]{\includegraphics[width=0.38\linewidth, keepaspectratio]{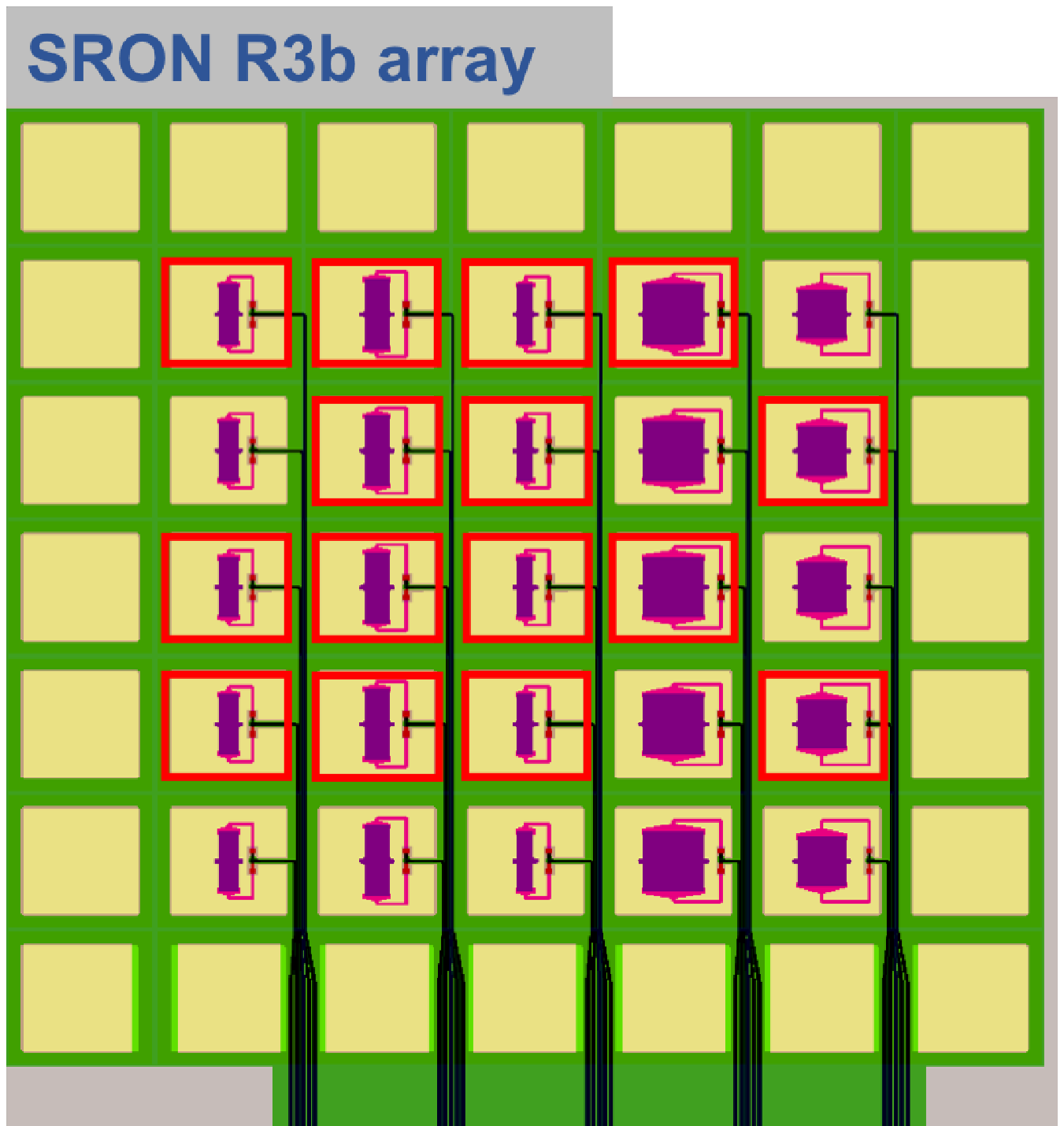}\label{fig:arrayR3b}}\hspace{0.1in}
	\subfloat[]{\includegraphics[width=0.4\linewidth, keepaspectratio]{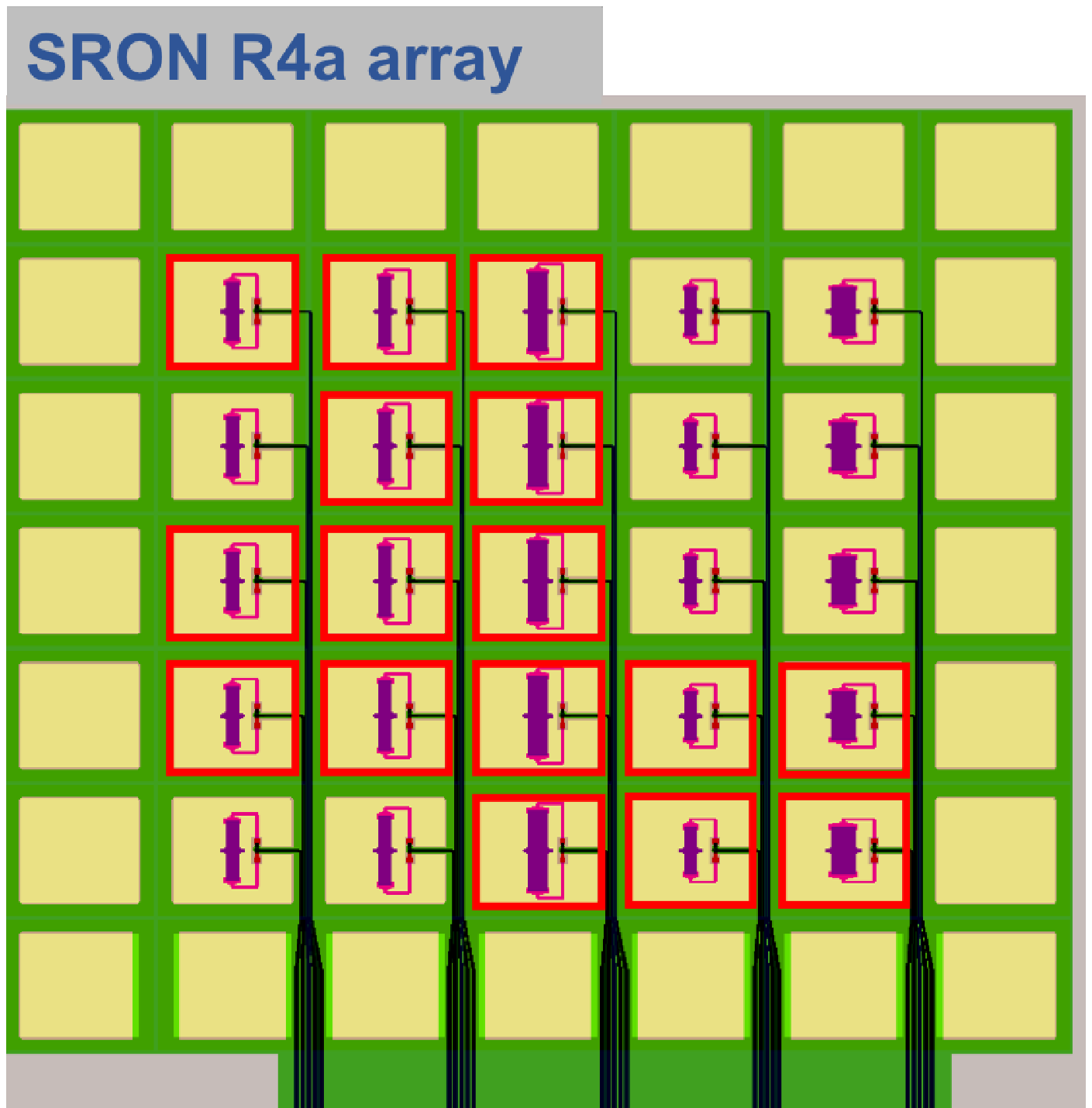}\label{fig:arrayR4a}}
\caption{5$\times$5 SRON mixed arrays. Array labelled R3b with high-aspect ratio devices with the following sizes (from left to right): 100$\times$30, 120$\times$40, 100$\times$25, 100$\times$100 and 80$\times$80 $\mu$m$^2$ ({\it Fig. ~\ref{fig:arrayR3b}}). Array labelled R4a with very high-aspect ration devices with the following sizes (from left to right): 100$\times$20, 120$\times$20, 140$\times$30, 80$\times$20 and 80$\times$40  $\mu$m$^2$ ({\it  Fig. ~\ref{fig:arrayR4a}}).  Each column has devices with the same aspect ratio. {\it Red squares} identify the TESs that were connected and measured. (Color figure online.)}
\label{fig:SRONarrays}
\end{center}
\end{figure}

\begin{figure}[htbp]
\begin{center}
	\includegraphics[width=0.7\linewidth, keepaspectratio]{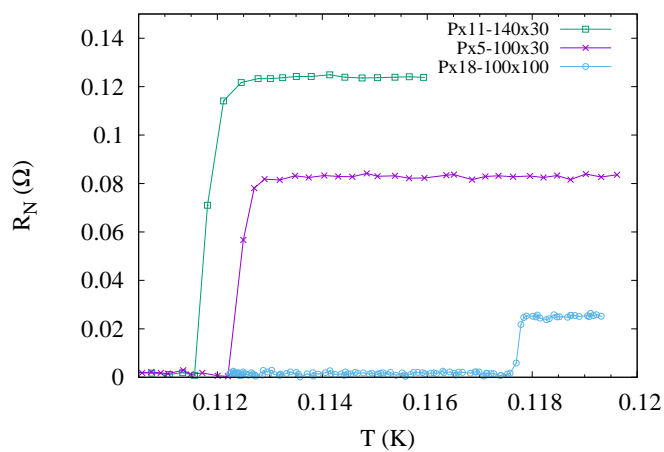}
\caption{4-terminal resistance measurement of pixels with three different aspect ratios: 100$\times$100 ({\it blue dots}), 100$\times$30 ({\it purple crosses}) and 140$\times$30  $\mu$m$^2$ ({\it green squares}) measured with the same excitation current of 3~$\mu$A. Lines serve no other purpose than to guide the eye. (Color figure online.)}
\label{fig:RvsT}
\end{center}
\end{figure}

Fig. ~\ref{fig:SRONarrays} shows the design of two different 5$\times$5 mixed arrays that are fabricated on a wafer together with other arrays. We can identify one array, labelled array R3b, as containing high-aspect ratio devices with the following dimensions (length$\times$width) (from left to right in Fig. ~\ref{fig:arrayR3b}): 100$\times$30, 120$\times$40, 100$\times$25, 100$\times$100 and 80$\times$80 $\mu$m$^2$, while the other array, labelled array R4a, contains very high-aspect ratio devices with the following sizes (from left to right in Fig. ~\ref{fig:arrayR4a}): 100$\times$20, 120$\times$20, 140$\times$30, 80$\times$20 and 80$\times$40  $\mu$m$^2$.  Note that every column has the same detector size in order to study the performance of the TES as a function of the bias frequency and to have reasonable statistics on pixel with the same design. The red squares point out the 31  TESs that have been connected and measured. All the TESs  have the same bilayer thickness of Ti (35 nm) and Au (200 nm) and use a 0.5 $\mu$m thick SiN membrane. We measured a normal squared resistance R = 25~m$\Omega$/sq and  a critical temperature T$_c\sim$115~mK. The thermal conductance G at T$_c$ ranges from 107 to 172~pW/K and from 61 to 104~pW/K for the different aspect-ratio arrays R3b and R4a, respectively.\\
Recently it has been shown\cite{ullom} that, depending on the quality of the Bismuth in the absorber, a non-Gaussian thermal response can be induced in the energy spectra. In order to identify the real performance of the pixel design, all the absorbers consist only of Au 2.3~$\mu$m thick and have the same size (240$\times$240 $\mu$m$^2$) with a heat capacity C = 1.1~pJ/K at T$_c$. Each absorber has, at its corners, four contacting points to the membrane  and two central contacting points directly to the sides of the TES. More details on fabrication of the SRON TES array will be published in this special issue \cite{ken}.\\
Due to the thicker bilayer used in this work, compared to the previous batch \cite{pourya}, we need to tailor the aspect ratio to obtain the required normal resistance and bias power. Fig.~\ref{fig:RvsT} shows the 4-terminal resistance measurements as a function of bath temperature measured for TESs with different dimensions which are a squared 100$\times$100, a 100$\times$30   and a 140$\times$30 $\mu$m$^2$ TES.\\
The characterization of the two arrays was performed in two different experimental measurement setups, named XFDMLarge for the R3b array and named XFDMProbe for the R4a array. Both setups were placed in the same dilution refrigerator that can provide a bath temperature of  $\sim $ 40~mK. TESs were characterised under AC bias using an existing FDM readout system (1--5~MHz) \cite{fdm} in the single-pixel mode configuration. Each TES is connected in series with an LC resonator on a LC filter chip with a coil inductance L = 2~$\mu$H and a 1:2 transformer chip for the XFDMLarge and an L = 1 $\mu$H and a direct connection to the TESs (no transformers) for the XFDMProbe. The TES array chips and the cryogenic components of FDM readout were mounted in a low magnetic impurity copper bracket fitted into an Al superconducting shield.
The bracket of each setup also accommodates a heater, a thermometer and a Helmholtz coil. The heater and thermometer are used to regulate the temperature locally on the chip. The coil is for applying a uniform magnetic field perpendicular to the TES array to compensate any remnant magnetic field trapped in the experiment setup.

\section{Energy Resolution}
\begin{figure}[htbp]
\begin{center}
	\subfloat[]{\includegraphics[width=0.4\linewidth, keepaspectratio]{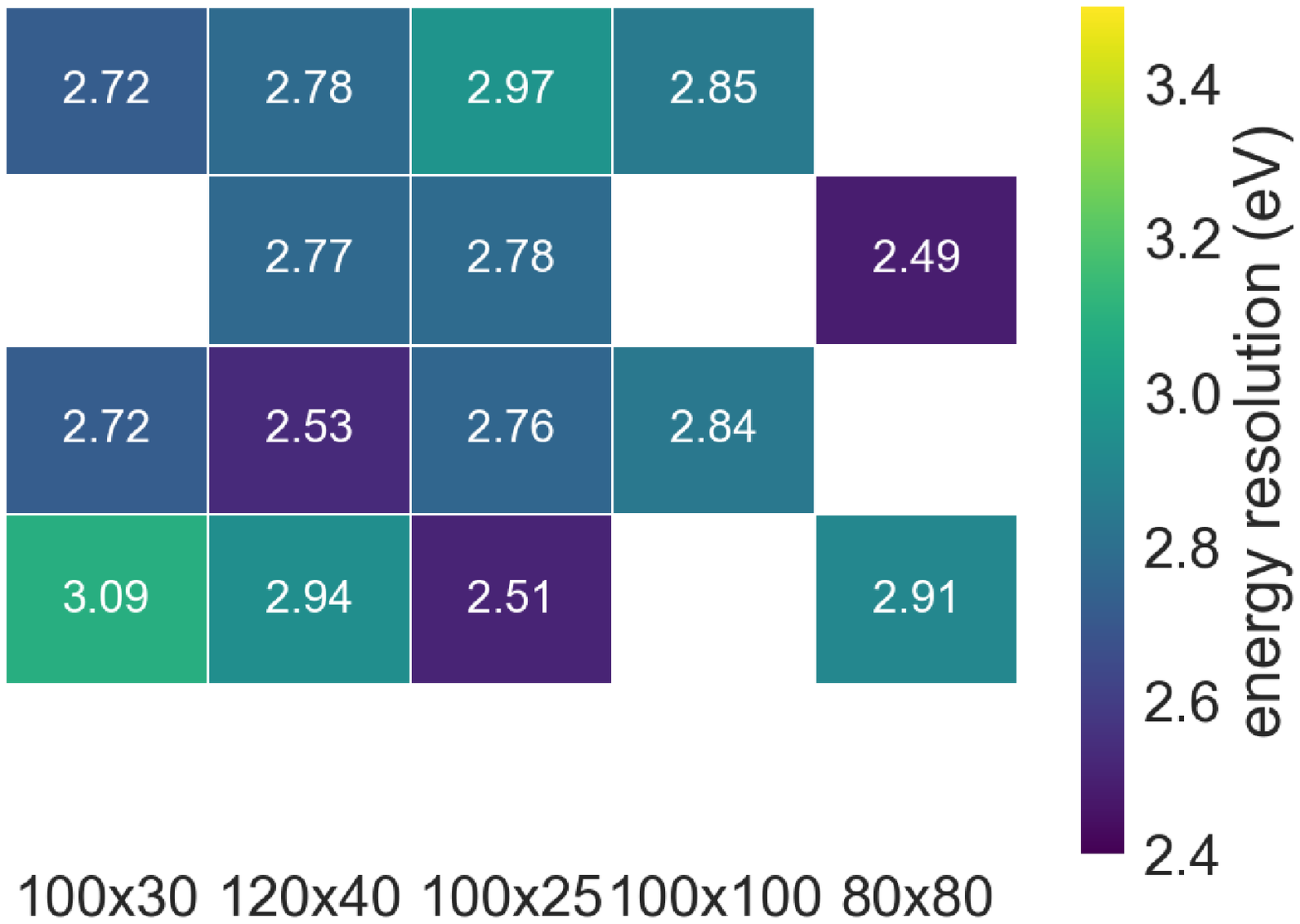}\label{fig:eVmapR3b}}\hspace{0.15in}
	\subfloat[]{\includegraphics[width=0.4\linewidth, keepaspectratio]{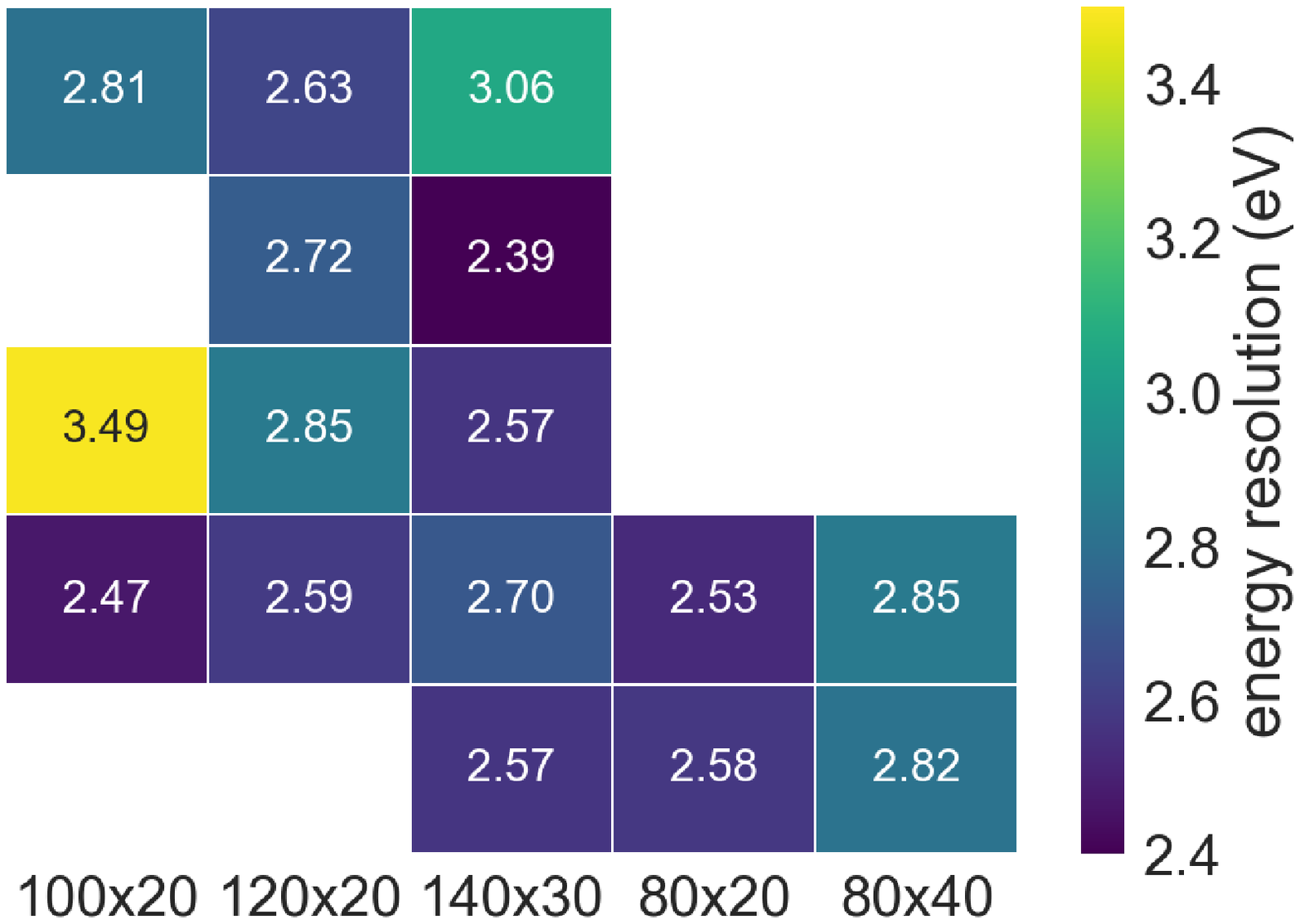}\label{fig:eVmapR4a}}
\caption{Overview of the-single pixel energy resolution measured in array R3b array ({\it Fig. ~\ref{fig:eVmapR3b}}) and in array R4a ({\it Fig. ~\ref{fig:eVmapR4a}}). Consider an error bar of $\pm$0.15 eV for each value. (Color figure online.)}
\label{fig:eVmap}
\end{center}
\end{figure}

\begin{figure}[!htbp]
\begin{center}

	\subfloat[TES14-140$\times$30 $\mu$m$^2$, f$_0$=1.6 MHz]{\includegraphics[width=0.45\linewidth]{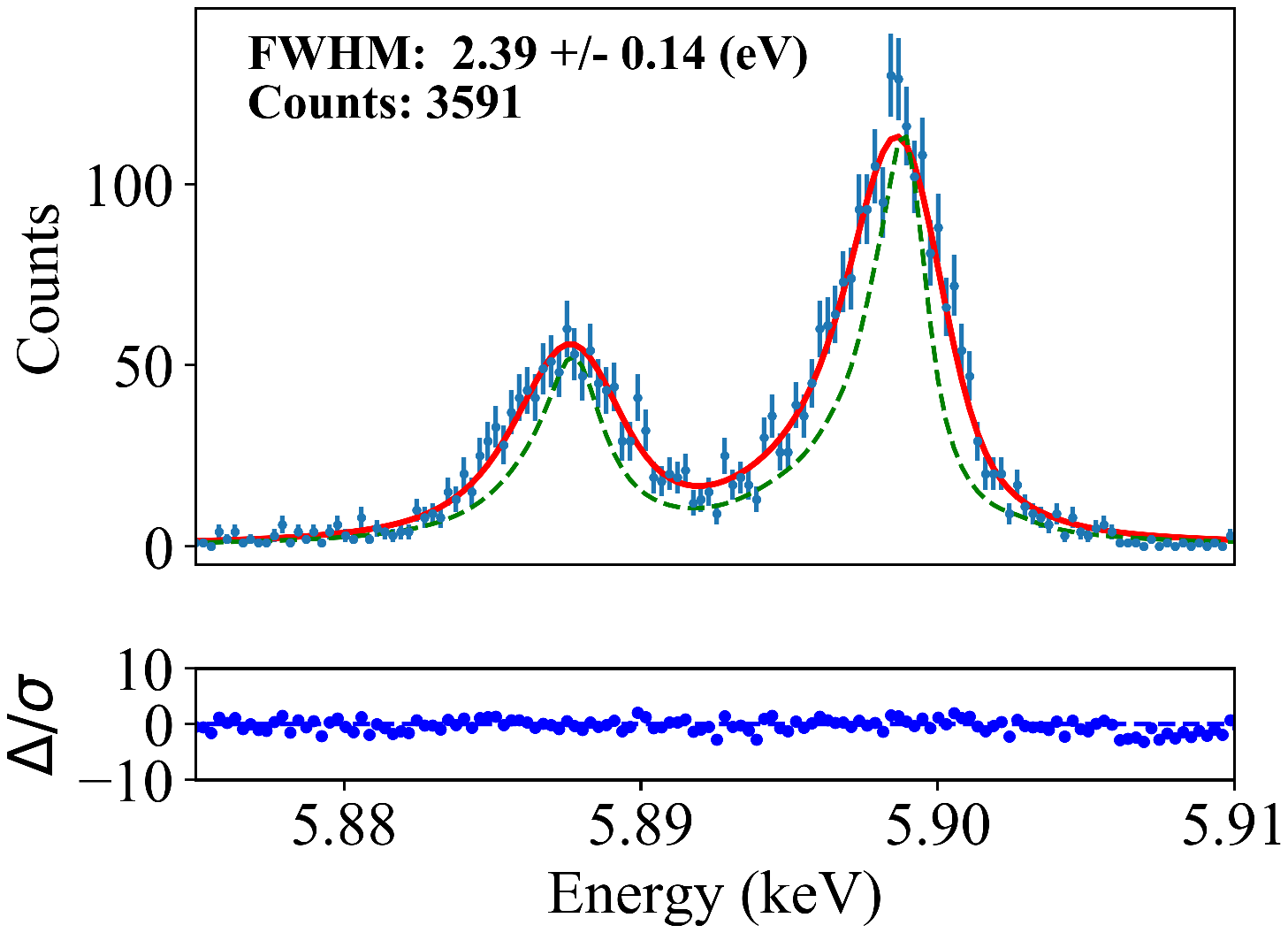}\label{fig:px6}}
 	\subfloat[TES24-80$\times$80 $\mu$m$^2$, f$_0$=2.9 MHz]{\includegraphics[width=0.45\linewidth]{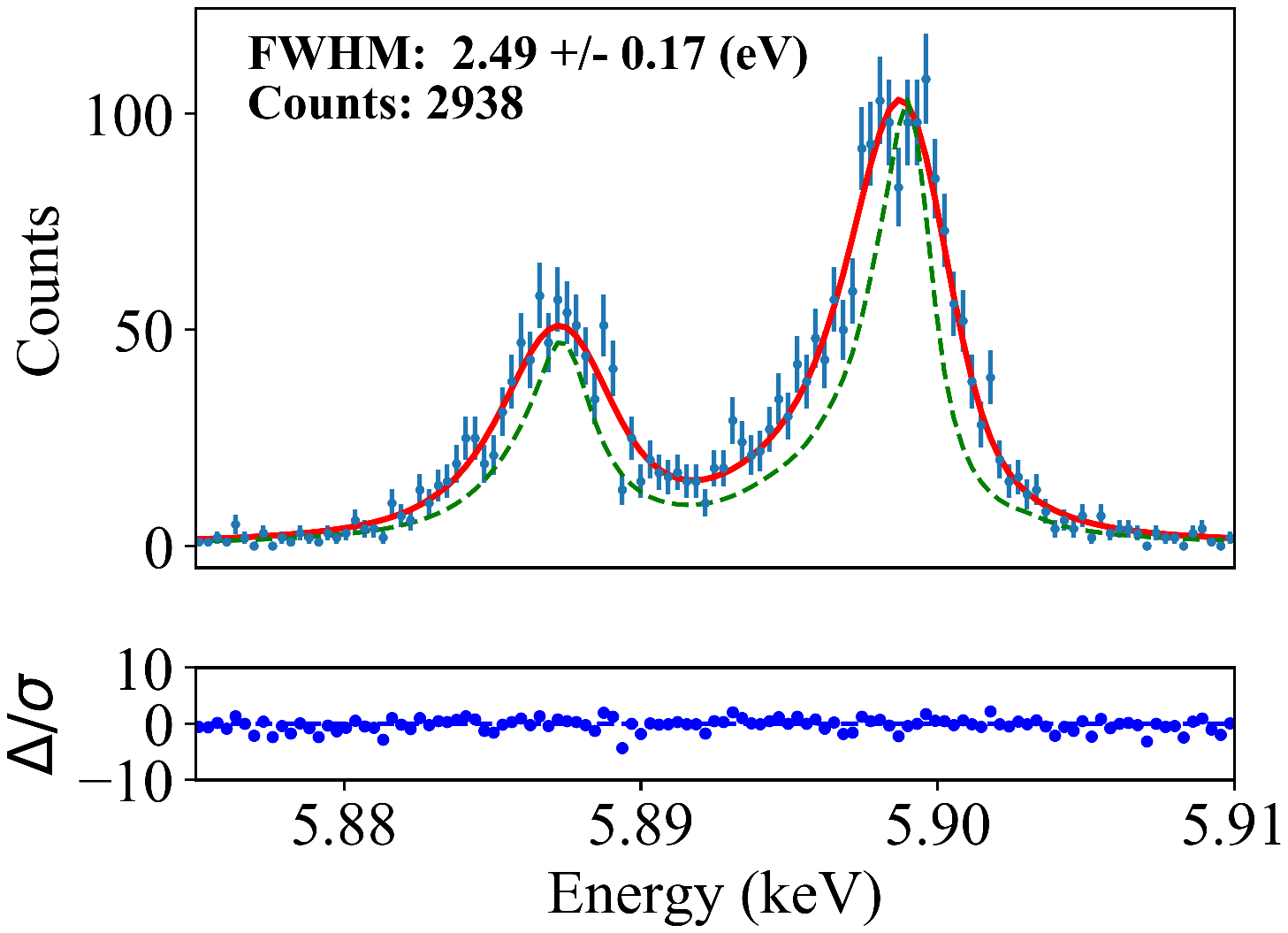}\label{fig:px11}}\\
	
	\subfloat[TES12-100$\times$25 $\mu$m$^2$, f$_0$=3.8 MHz]{\includegraphics[width=0.45\linewidth]{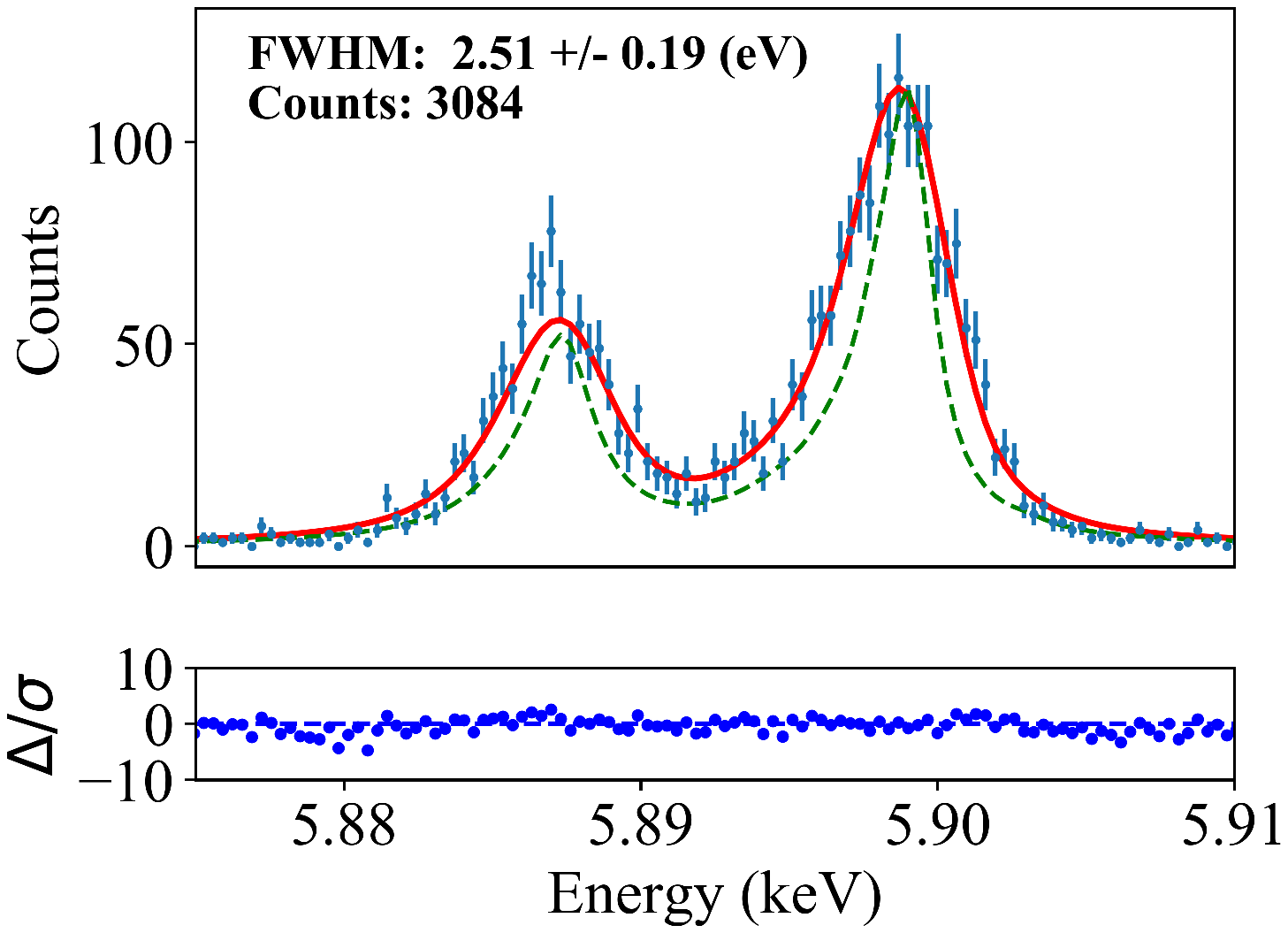}\label{fig:px13}}
	\subfloat[TES13-100$\times$25 $\mu$m$^2$, f$_0$=4.6 MHz]{\includegraphics[width=0.45\linewidth]{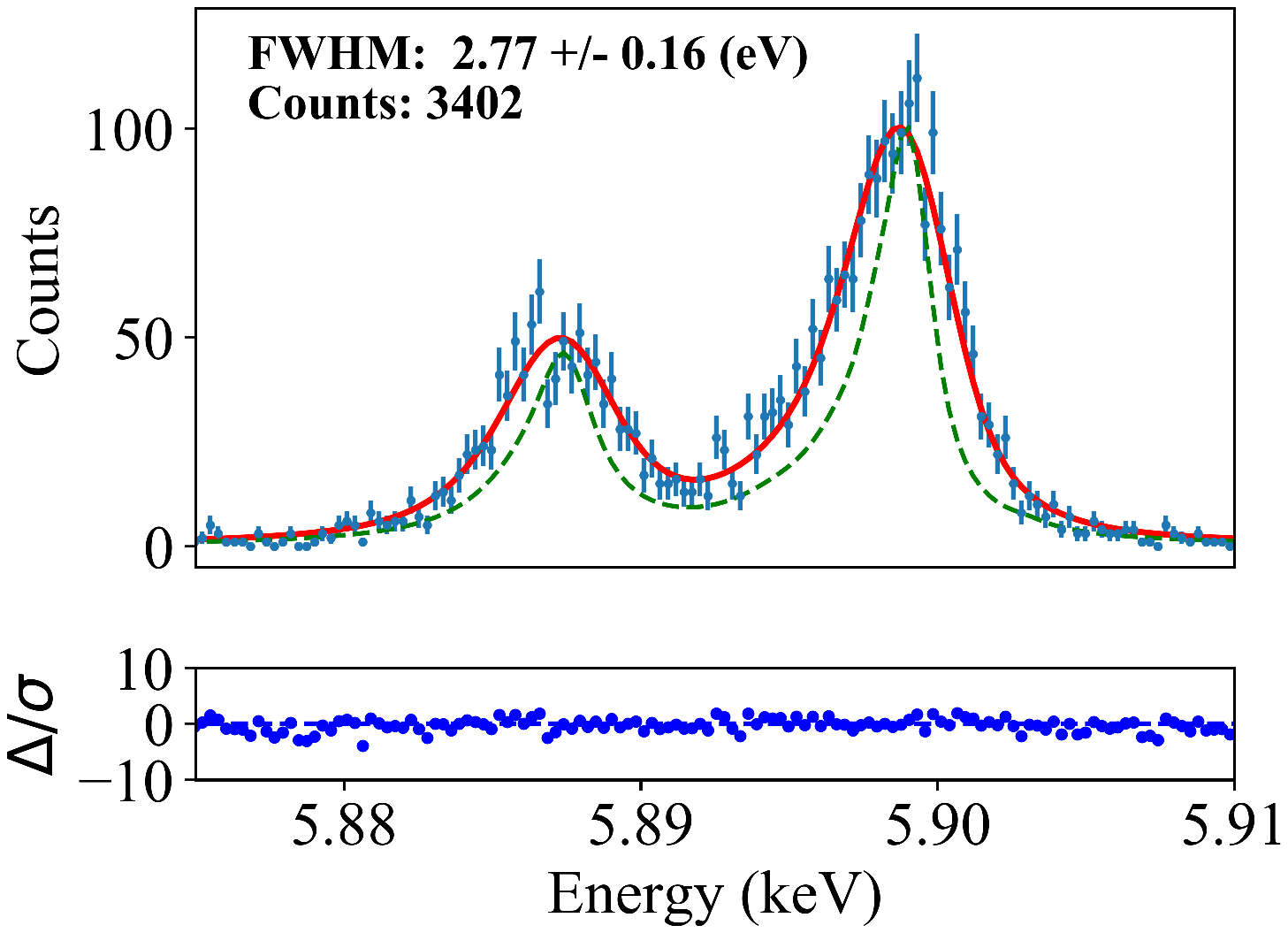}\label{fig:px16}}\\

\caption{Small selection of the best spectra at 5.9 keV for TESs mounted in the XFDMLarge setup (from Fig. ~\ref{fig:px11} to \ref{fig:px16}) and in the XFDMProbe  (Fig. ~\ref{fig:px6}) with bias points between 15-20\% of R$_N$. The {\it red solid line} is the best fit to the data, and the {\it blue points} are the measured Mn-K$\alpha$ emission  lines. The lower plots show the residuals of the fit normalised by the error bars. (Color figure online)}
\label{fig:eV}
\end{center}
\end{figure}
\begin{figure}[!htbp]
\begin{center}
	\subfloat[]{\includegraphics[width=0.45\linewidth]{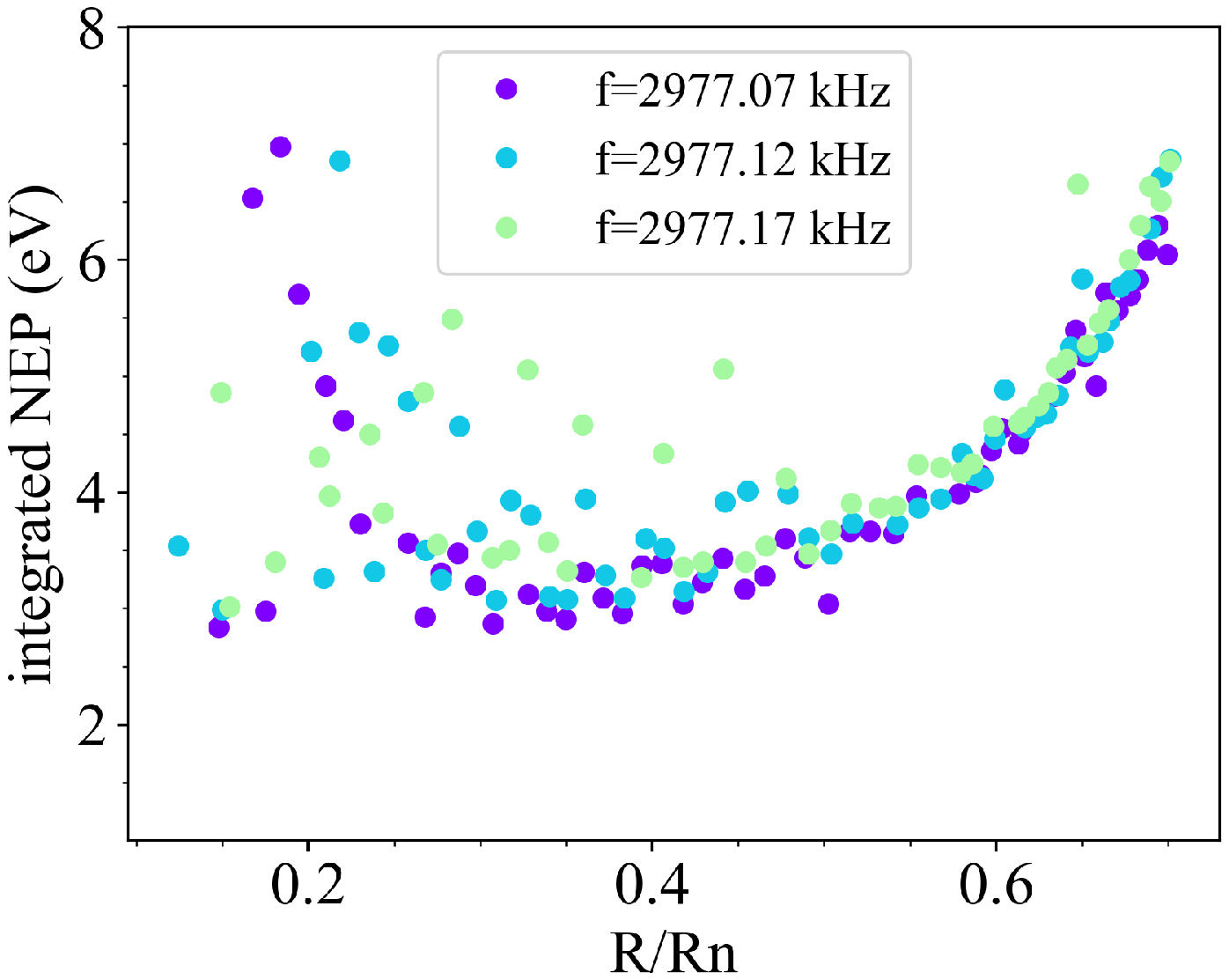}\label{fig:neppx11}}
	\subfloat[]{\includegraphics[width=0.45\linewidth]{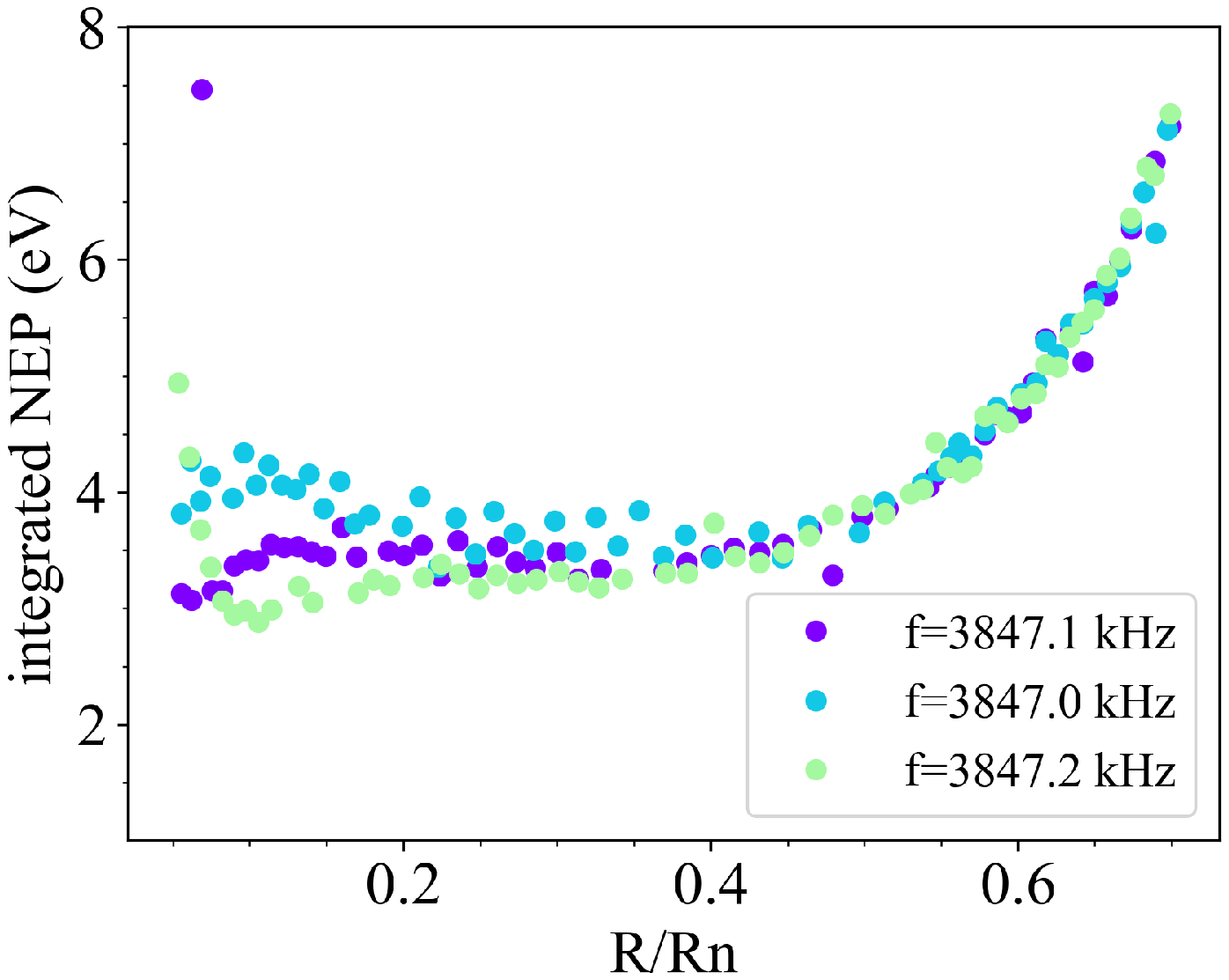}\label{fig:neppx13}}
\caption{Noise Equivalent Power (NEP) scan for TES24-80$\times$80 $\mu$m$^2$ ({\it Fig. ~\ref{fig:neppx11}}) and for TES12-100$\times$25 $\mu$m$^2$ ({\it Fig. ~\ref{fig:neppx13}}) as a function of bias point for three different bias frequencies.}
\label{fig:nep}
\end{center}
\end{figure}

An Fe-55 X-ray source is placed closely above the array in both the setups and illuminates the entire TES array at a count rate of $\sim$1 cps per pixel for the given absorber with the Mn-K$\alpha$ 5.9-keV fluorescent X-ray line.  We measured the X-ray energy resolution for each TES connected in the bias range of 10--20\% of R$_N$,  where the Noise Equivalent Power (NEP) scan usually predicted the lowest energy resolution. Fig. ~\ref{fig:eVmapR3b} and ~\ref{fig:eVmapR4a} show a graphical impression of the global energy resolution measured for arrays R3b and R4a, respectively. About 20 of the pixels show an excellent energy resolution between 2.4 and 2.8~eV, eight of them between 2.8 and 3~eV and only three are worse than 3~eV.\\
In Fig. ~\ref{fig:eV} we present four typical spectra, as examples. The best energy resolution obtained for array R3b mounted in the XFDMLarge setup is shown in Fig. ~\ref{fig:px11}. It is worth noting how the energy resolution in both the setups remains excellent when devices with the same aspect ratio are biased at different resonant frequency f$_0$ as shown for instance in Fig. ~\ref{fig:px13}-\ref{fig:px16}. The  best energy resolution 2.39$\pm$0.14~eV at 5.9~keV was measured in the XFDMProbe for the high-aspect ratio 140$\times$30 $\mu$m$^2$ biased at a resonant frequency of 1.6 MHz and between 10-20\% of R$_N$.   This result is shown in Fig. ~\ref{fig:px6}. 
At this stage we are not yet able to identify the best TES design illustrating the ultimate energy resolution. It is also still unclear, whether there is a correlation between the energy resolution and the different aspect ratios of the TES. In the ideal scenario, we would expect to find the identical energy resolution along the columns of Fig. ~\ref{fig:eVmap} and possibly a significant variation between them, according to the  pixel design variation. However, we still have to deal with the different behaviour of the TES at the different bias frequencies, which is reflected in the spread of the energy resolution along a single column. If we consider the R3b array (Fig. ~\ref{fig:eVmapR3b}) and take the mean of the energy resolution for each column, we get a uniform energy resolution ranging from 2.7 to 2.84~eV, although the different squared TESs (last two columns) have worse statistics, i.e. small spread in bias frequencies.  In the R4a array these averages are slightly more scattered,  giving an energy resolution ranging from 2.55 up to 2.86~eV. This could be due to the broader variation in the pixel design but also to the poorer statistics for the last two columns of Fig. ~\ref{fig:eVmapR4a}. Of course, with the upcoming measurements on other high aspect-ratio mixed arrays and on the uniform kilo-pixel array, we might define the relationship between the detector energy resolution, the pixel design and the bias frequency. What is clear is that seven pixels out of 16 in R4a (mounted in the XFDMProbe setup) show an energy resolution around 2.5~eV, which is a slightly better than that found for array R3b (3 out of 15). This difference is likely due to the better temperature stability ($\sim$ 1.9~$\mu$K at 55~mK) in the XFDMProbe; the XFDMLarge ($\sim$ 3~$\mu$K at 50~mK) is less stable. Furthermore, we notice that square or close-to-square TESs with a lower normal resistance are more affected by the weak-link effect, showing an oscillation in the predicted energy resolution as can be seen in the NEP plotted in Fig. ~\ref{fig:neppx11}. This means that, although there might be very good energy resolution (Fig. ~\ref{fig:px11}), this happens only for a limited number of bias points through the transition. On the other hand, high aspect-ratio devices with a higher normal resistance show a flat NEP over a wide bias range between 5 and 30\% of R$_N$ as shown in Fig. ~\ref{fig:neppx13}.\\
It is important to highlight that these results have been obtained with a  T$_c$ of 110~mK and for these arrays this could be a limiting factor. Reducing the T$_c$ down to $\sim$ 90~mK should give an improvement in the energy resolution of between 15 and 20\%, approaching even closer to the X-IFU requirements (2.5~eV below 7~keV over the whole array).

\vspace{-0.5cm}
\section{Conclusion}

New  5$\times$5 TiAu  TES mixed arrays, containing TESs with a thick bilayer, but varying  aspect-ratios  of  the TES, were fabricated and experimentally evaluated with 5.9-keV X-ray photons.  Those TESs have a T$_c$ of 110 mK, no normal metal structures on the top of the bilayer,  and make use of  a thinner SiN membrane (0.5~$\mu$m). 
We measured  an energy resolution between 2.4 and 2.8 eV at 5.9 keV on 20 out of the 31 TESs. New TESs with  a T$_c$ $\sim$ 90 mK should  improve  the energy resolution further  and meet the X-IFU requirements for all the pixels in the array.\\
Our microcalorimeters  represent the best TES microcalorimeters ever reported in Europe. Not only are our detectors  able to meet the detector requirements of the X-IFU instrument, but they also potentially offer technology for other future X-ray space missions, fundamental physics experiments, plasma characterization,  and material analysis. We are now ready to test a kilo-pixel array in combination with the FDM readout in multi-pixel mode.\\

\begin{acknowledgements}
This work is partly funded by European Space Agency (ESA) and coordinated with other European efforts under ESA CTP contract ITT AO/1-7947/14/NL/BW. It has also received funding from the European Union's Horizon 2020 Programme under the AHEAD (Activities for the High-Energy Astrophysics Domain) project with grant agreement number 654215.
\end{acknowledgements}


\end{document}